\documentclass[nature,iicol]{sn-jnl}

\usepackage{graphicx}%
\usepackage{multirow}%
\usepackage{amsmath,amssymb,amsfonts}%
\usepackage{amsthm}%
\usepackage{mathrsfs}%
\usepackage[title]{appendix}%
\usepackage{xcolor}%
\usepackage{textcomp}%
\usepackage{manyfoot}%
\usepackage{booktabs}%
\usepackage{algorithm}%
\usepackage{algorithmicx}%
\usepackage{algpseudocode}%
\usepackage{listings}%
\usepackage{color}

\theoremstyle{thmstyleone}%
\theoremstyle{thmstyletwo}%

\theoremstyle{thmstylethree}%

\begin{document}

\title{Quantum Entanglement in Nuclear Fission}


\author[1]{\fnm{Yu} \sur{Qiang}}

\author*[1,2]{\fnm{Junchen} \sur{Pei}}\email{peij@pku.edu.cn}

\author[3]{\fnm{Kyle} \sur{Godbey}}

\affil*[1]{\orgdiv{State Key Laboratory of Nuclear Physics and Technology, School of Physics}, \orgname{Peking University}, \orgaddress{ \city{Beijing}, \postcode{100871}, \country{China}}}

\affil[2]{\orgdiv{Southern Center for Nuclear-Science Theory (SCNT)}, \orgname{Institute of Modern Physics, Chinese Academy of Sciences},\orgaddress{ \city{Huizhou}, \postcode{516000},  \country{China}}}

\affil[3]{\orgdiv{Facility for Rare Isotope Beams},\orgname{Michigan State University},\orgaddress{East Lansing, Michigan},\postcode{48824},
\country{USA}}


\abstract{
Nuclear fission presents a unique example of quantum entanglement in strongly interacting many-body systems.
A heavy nucleus can split into hundreds of combinations of two complementary fragments in the fission process.
The entanglement of fragment wave functions is persistent even after separation and 
impacts the partition of particles and energies between fragments.
Based on microscopic dynamical calculations of the fission of $^{240}$Pu, 
this work finds that  quantum entanglement is indispensable in the appearance of sawtooth distributions of average excitation energies of fragments
and thus neutron multiplicities,
but not in average neuron excess of fragments. Both sawtooth slopes from particle-number projections are found to be steep -- a feature which can be alleviated by random fluctuations.
These findings may impact the understanding of quantum entanglement more broadly in
mesoscopic systems.  
}




\maketitle

Quantum entanglement has been well testified in paired two photons and it becomes far more complicated in
many-body systems~\cite{manyb}. In the nuclear fission process, quantum entanglement is expected to be present between two complementary fragments~\cite{future} and this entanglement of fragment wave functions is persistent after the scission process~\cite{youngs}.
Despite this expectation, the extent to how quantum entanglement can play a role in fission observables is not known.
Among existing fission data, the sawtooth pattern is still a puzzling phenomena and has been exhibited in distributions of multiple average fission observables such as neutron excess ($\langle N \rangle /Z$ ratios) in dependence of fragment charge~\cite{nzratio1,nzratio2,nzratio3},
neutron multiplicities~\cite{neutron1,neutron2}, $\gamma$-ray multiplicities~\cite{gamma1,gamma2} and recently the angular momentum in dependence of fragment mass~\cite{wilson}.
These sawtooth structures provide a unique probe of the sharing mechanism of particle numbers and excitation energies, i.e. the quantum entanglement effect, between two nascent fragments.

The sharing of particles and energies is a result of complex  non-equilibrium and non-adiabatic dynamical evolutions.
Both quantum effects and statistical fluctuations play a role at low energy fission.
The sawtooth structure was explained by the
shape-dependent level densities of fragments based on Metropolis random walks on multi-dimensional potential energy surfaces (PES)~\cite{neutron3}.
This statistical sharing mechanism, however, is not a complete description as light fragments generally have less excitation energies than heavy fragments~\cite{neutron3},
being inconsistent with experiments at low energies.
The Langevin model~\cite{langevin1,langevin2} and time-dependent generate-coordinate-method (TD-GCM)~\cite{tdgcm}
can reasonably describe fission yields based on a static PES.
Note that random walks, the Langevin model, and TD-GCM don't involve the splitting dynamics and are incompatible with self-consistent energy sharing mechanisms.
In this respect, a promising framework to understand sawtooth structures in fission observables is the microscopic time-dependent density functional theory (TD-DFT).
In TD-DFT the sharing of particles and excitation energies, shapes of nascent fragments, total kinetic energies, energy dependence, equilibration effects, non-adiabatic dissipative effects, and scission configurations can be described self-consistently~\cite{koonin,negele,tddft,simenel2020,stevenson,bulgacprl,bulgacprc,scampsnature,formation,ayikprl,qy1,qy2}.

Enabled by advancements in high performance computing, TD-DFT has brought fruitful insights in nuclear fission and reactions in recent years.
A longstanding issue of TD-DFT is the outcome distributions of fission observables are too narrow~\cite{bulgacprc,stevenson}, indicating fluctuations are underrepresented.
Previously, TD-DFT with additional initial fluctuations~\cite{ayikprl,lacroix2014} or dynamical fluctuations~\cite{bulgac-fluc,qy1}  have been proposed
to produce the distributions of fission yields.
Another issue in TD-DFT is the symmetry breaking at the mean-field level.
To this end, particle number projections (PNP)~\cite{sekizawa,258fm,bulgacproj} and angular momentum projections ~\cite{bulgacspin}
are adopted to restore symmetries and also build quantum correlations beyond TD-DFT.
It has been demonstrated that PNP can obtain distribution widths of fission fragment yields~\cite{schunckproj}.
These remedies to TD-DFT are also widely used in studies of heavy-ion reactions for similar motivations~\cite{sekizawa,godbey2020}.

Here we study the sharing of particles
and energies between fragments and the role of quantum entanglement in sawtooth structures, by combining random fluctuations and
PNP based on TD-DFT.
Compared to PNP in the literature~\cite{sekizawa,258fm,bulgacproj,schunckproj},
it has to be noted that we implement a double projection scheme to incorporate the
 couplings between two fragments.
Furthermore,  we have performed calculations of fragment energies after projections
to study energy sharing.  In fact, the average  neutron multiplicities are determined by excitation energies of fragments~\cite{nzratio3,bulgacprc}.
The projection procedure is by analogy a measurement to decode the entangled fragments.
The reduced entanglement entropy between two fragments is calculated to estimate the entanglement magnitude.


\section*{Results}

\begin{figure}[t]
\centering
\includegraphics[width=0.46\textwidth]{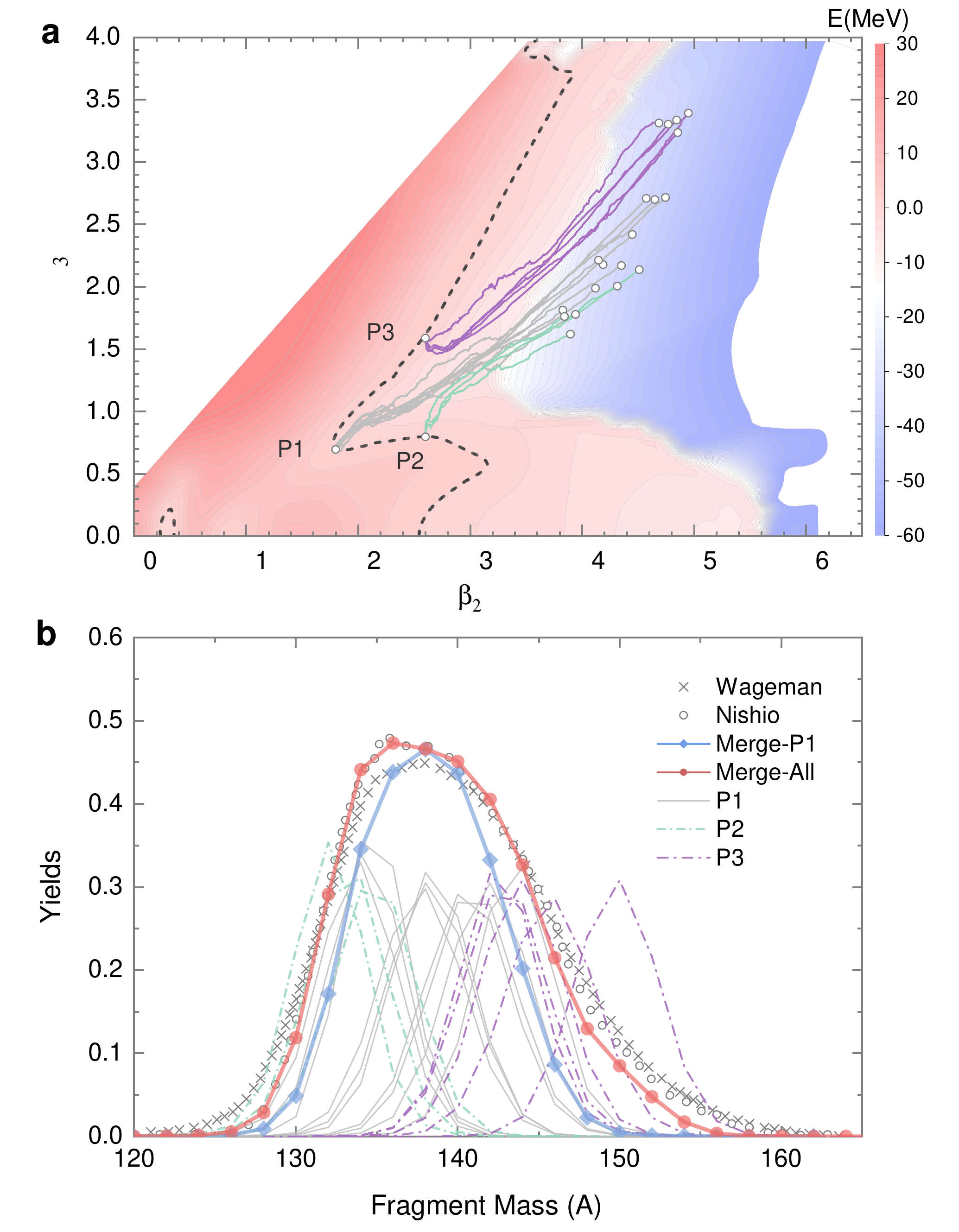}
\caption{ {\bf Calculated primary fission mass yields of the heavy fragment from the fission of $^{240}$Pu.}
(a) trajectories of fluctuated fission events  in the PES of quadrupole-octupole deformations ($\beta_2$, $\beta_3$).  
 (b) the mass yields obtained by double PNP on fluctuated events from TD-BCS evolutions.
The blue line denotes the merger of events from P1 and the red line denotes the merger of all events.
The experimental mass yields are taken from ~\cite{wageman,nishio}.
                             \label{FIG1}
}
\end{figure}

\noindent{\bf{Fission Mass Yields}}\\
Firstly the fragment mass yields of $^{240}$Pu are calculated by combing
random fluctuations and double PNP based on TD-DFT, as shown in Fig.\ref{FIG1}.
The initial deformation in TD-BCS evolutions at zero temperature is chosen from the two-dimensional PES as shown in Fig.\ref{FIG1}(a).
With double PNP, the mass distributions of light and heavy fragments are symmetric and only the distribution of heavy fragments is
shown in Fig.\ref{FIG1}(b).
The spreading fission events with different scission deformations are obtained, which are associated with fluctuated trajectories.
The fission events are mainly taken from P1 point, which are the most probable events. 
The projection on each event
results in a narrow distribution of mass yields, for which the width is about 8 mass units. This is consistent with earlier projection calculations~\cite{schunckproj}.  The merger of events from P1 produces a considerable distribution of
mass yields around the peak. 
However, only the merger of all events from P1, P2 and P3 on the same contour can reproduce the distribution of mass yields.
This is because the employed random fluctuations are not sufficient, and events from P2 and P3 are added to study additional initial fluctuations perturbatively.
The results could be slightly dependent on the choice of P2 and P3, nevertheless, 
the  combined fission events merely provide a basis for following studies of the sharing mechanism, rather than reproduce fission yields in this work.

\begin{figure*}[t]
\centering
\includegraphics[width=0.9\textwidth]{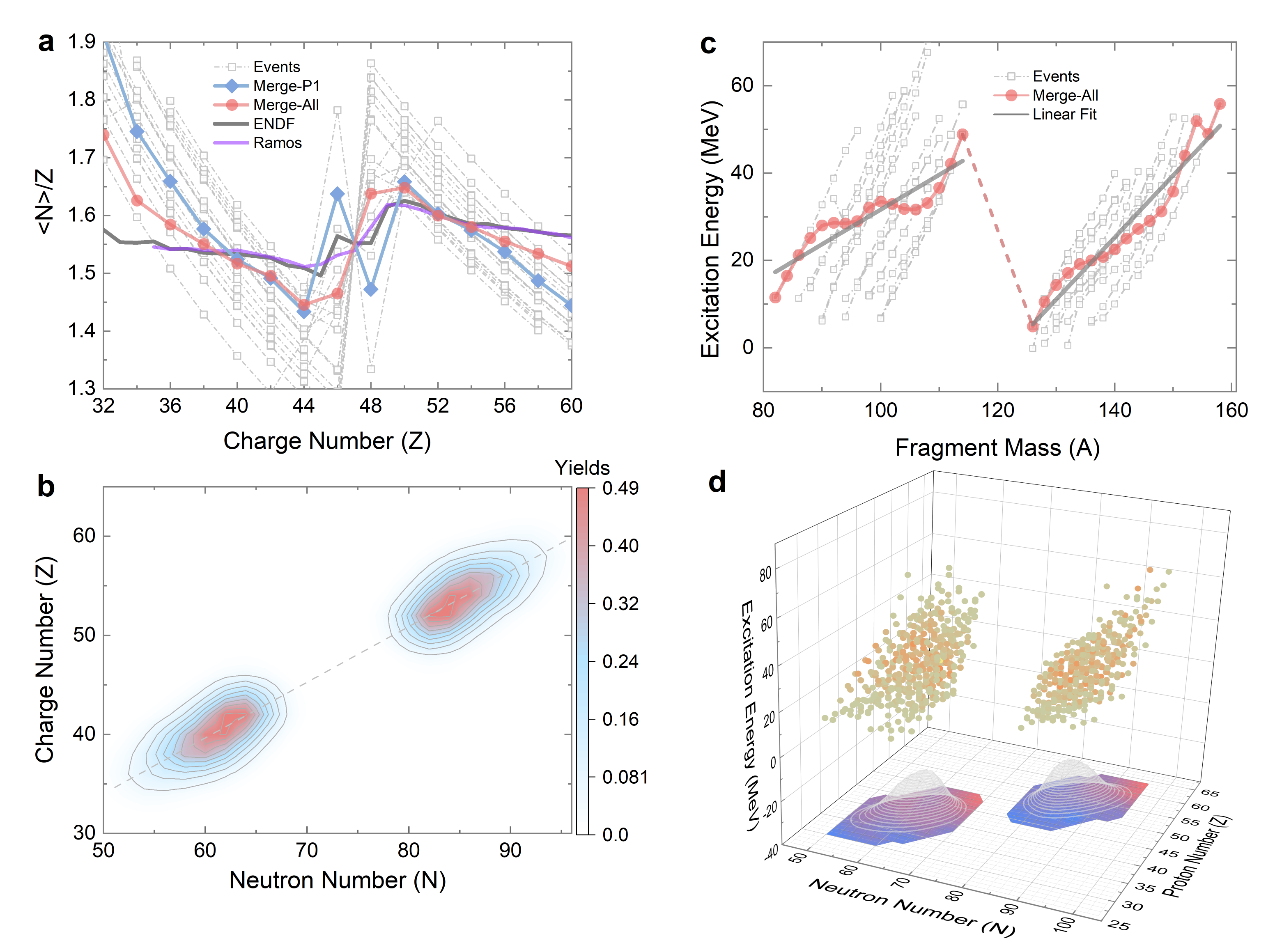}
\caption{
{\bf Sawtooth structures in $\langle N \rangle /Z$ ratios and excitations energies of fragments. }
(a) the gray-dashed  lines denote $\langle N \rangle /Z$ ratios in fragments from each fluctuated fission event by double PNP and the red line denotes merged results of all events in the
fission of $^{240}$Pu.
The ENDF evaluation data~\cite{ENDF} (black line) and experimental data from D. Ramos {\it et al.}~\cite{nzratio2} (purple line) are also shown,
in which IFY data are shifted  by considering the emission of two neutrons
 so that different data are converged at $Z$=47.
(b) calculated two dimensional fission yields corresponding to the merger of all events.
(c) the gray-dashed lines show excitation energies from projection calculations of each fluctuated fission event.
The merger of all events is denoted by the red line. 
(d) the excitation energies of all projected fragments are displayed as a three-dimensional view,
in which the same fragment is produced by different events with different projection probabilities (denoted by the point colour). 
The distribution of mass yields is constructed by contour lines. The 2D view of excitation energies is illustrated in the $N$-$Z$ plane.
 \label{FIG2}
}
\end{figure*}

\noindent{\bf{Sawtooth Structures}}\\
In recent years, accurate measurements of complete isotopic fission yields are realized~\cite{nzratio2,nzratio3}, providing
 valuable data for comprehensive understandings of nuclear fission.
 In particular, the average neutron excess $\langle N \rangle /Z$ of fragments in dependence of charge numbers
 shows inverted sawtooth structures~\cite{nzratio2}. 
 In Fig.\ref{FIG2}(a), $\langle N \rangle /Z$ ratios from
 projected fragments of each fluctuated event are shown.
The same fragment can come from projections on different events.
 The merger of all events generate a remarkable inverted sawtooth structure.
 In the merger,  different probabilities of each fragment are considered,  which are actually the expectation values of the projection operator.
 The slopes of all projected events are similar and negative, but such slopes are very steep.
 It is extraordinary that the slopes are reduced by the merger of all events.
 The $\langle N \rangle /Z$ ratios from evaluated independent fission yields (IFY) in ENDF~\cite{ENDF} and experimental data from Ref.\cite{nzratio2} are also shown for comparison.
 It can be seen that the realistic sawtooth structures are less significant.
The $\langle N \rangle /Z$ ratios from ENDF  have dents at $Z$=46 and 48.
Correspondingly, some events have similar dents at $Z$=46 and 48 due to contributions from binary fragments.
This is presented in ENDF data at thermal neutron energy~\cite{ENDF}, but not seen  experimentally at a higher excitation of 10.7 MeV~\cite{nzratio2}.
Without contributions of events from P2 and P3  as additional initial fluctuations,  the slopes at two sides are too large, and the dents in the middle are so significant.
This indicates that stronger fluctuations are needed to alleviate the sawtooth's slopes.

The calculated two-dimensional distribution of fission yields  is displayed in Fig.\ref{FIG2}(b).
The projection of each event leads to a near-round Gaussian distribution.
The merger of all events result in an elliptic-like 2D distribution,
while evaluated IFY corresponds to a rather elongated elliptic distribution~\cite{ENDF}.
The steep slope can be traced back to the wide elliptic distribution  in the isobaric direction.
TD-DFT calculations are expected to be improved by
obtaining more spreading widths from fluctuations.
In TD-GCM calculations plus double PNP, the obtained sawtooth structure is similarly very steep and the 2D distribution is also not so elongated as IFY~\cite{schunckp2}.
Then TD-GCM calculations need an additional convolution with a Gaussian function~\cite{schunckp2} or additional dissipation effects~\cite{zhaojie}.
Although the fission mass yields can be reasonably reproduced in Fig.\ref{FIG1},
the 2D distributions of fission yields provide more stringent examinations.

The sawtooth structures are also present in neutron emission multiplicities in dependence of fragment mass of actinide nuclei~\cite{neutron1,neutron2}.
This serves as a direct probe of energy sharing between two fragments as in TD-DFT the average excitation energies of the most probable paired fragments can be obtained which shows that light fragment has more
excitation energy than heavy fragment~\cite{qy1}.
As a further step, Fig.\ref{FIG2}(c) shows the distribution of  excitation energies of fragments from projection calculations of each event.
The slopes of these events are similar -- all are positive and steep.
The merger of all events generates a prominent sawtooth structure. This situation is similar to the inverted sawtooth structure in $\langle N \rangle /Z$ ratios.
Again we see that the merger of random fluctuated events can alleviate the slope.
The sawtooth structures are usually explained based on the shapes and shell structures of nascent fragments without invoking the splitting dynamics~\cite{neutron3,schunck-rev}.
This work provides an alternative explanation for sawtooth structures in $\langle N \rangle /Z$ ratios and neutron multiplicities (see also Fig.S1 in the Supplement),
based on the combination of double PNP and dynamical fluctuations in TD-DFT.

The excitation energies of all projected fragments are shown as a three-dimensional distribution in Fig.\ref{FIG2}(d).
This actually illustrates an event-by-event fission simulator, in which correlations between multiple fission observable are inherently included.
The same fragment can be produced by several projected events and thus have different excitation energies, so that  correlated distributions of fission observables rather than only average observables can be provided.
There are several existing well-developed fission simulators with cascade decays  for practical applications such as CGMF~\cite{CGMF}, FREYA~\cite{FREYA} and FIFRELIN~\cite{FIFRELIN}.
This work illustrates a potential microscopic fission simulator although more improvements are needed to approach the feature-set of the previously mentioned software.

\begin{figure*}[t]
\centering
\includegraphics[width=0.98\textwidth]{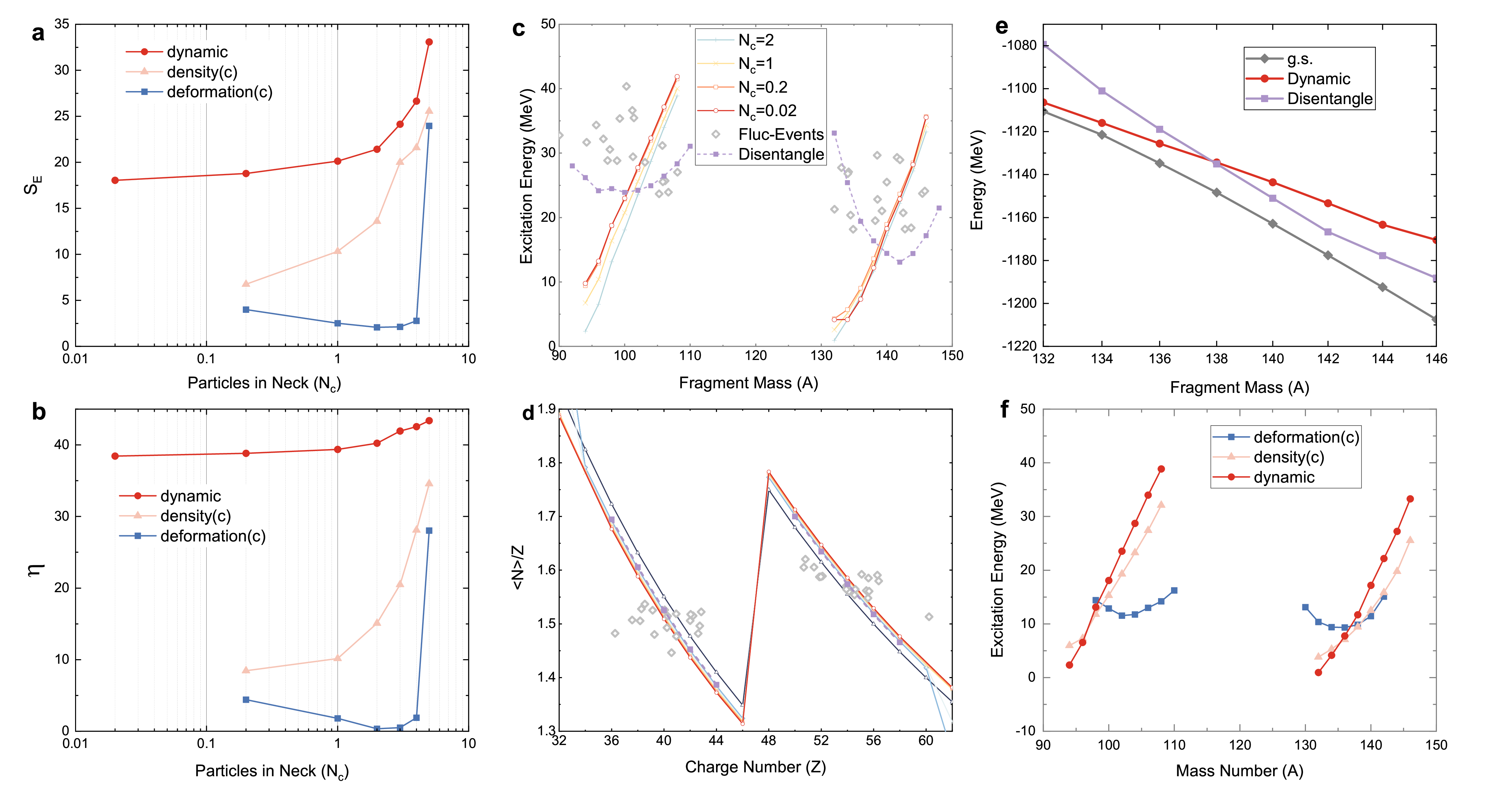}
\caption{
{\bf The quantum entanglement and the sawtooth slopes.}
The projection results of a single event around the peak of fission mass yields of $^{240}$Pu.
(a)the entanglement entropy $S_E$ is calculated as a function of $N_c$, in cases of non-adiabatic dynamical evolutions,  static density-constrained calculations of splitting fragments,
and static deformation-constrained calculations, respectively. 
(b) the approximate estimation of non-localization of wave functions between two fragments, $\eta$, as a function of $N_c$.
(c) calculated excitation energies with varying neck size $N_c$.
The scattered-spots denote random fluctuated events without projections.
Projection results of disentangled fragments at $N_c$=1.0 by separated density-constrained calculations are also shown for comparison.
(d) similar to (c) but for $\langle N \rangle /Z$ ratios.
(e) the projected binding energies of heavy fragments as a function of mass number, in cases of dynamical splitting, disentangled fragments and ground states, respectively. 
(f) the slopes of excitation energies from projection calculations, corresponding to different cases in (a) at  $N_c$=1.0. }
\label{FIG3}
\end{figure*}

\noindent{\bf{Quantum Entanglement}} \\
To further understand the sharing of particles and energies between two fragments, the quantum entanglement is studied. 
In the final splitting stage of a fission event around the peak, the entanglement entropy $S_E$ is calculated, as shown in  Fig.\ref{FIG3}(a), to estimate the entanglement magnitude as described in {\textsf{Methods}}, according to Ref.~\cite{entropy}.
$S_E$ is shown as a function of varying neck size, i.e., the number of particles in the neck $N_c$,
as an indicator of the separation process shown in Fig.S2 and Fig.S3 in the Supplement.
It can be seen that the dynamical entanglement decreases until the separation.
After scission, the entanglement is almost unchanging.
The straightforward estimation of non-localization of wave functions in two fragments is shown in Fig.\ref{FIG3}(b), as described in {\textsf{Methods}}, which is not strict but shows similar trends as the entanglement 
entropy towards scission. 

The persistent entanglement is mainly due to non-adiabatic dynamics since the final splitting  is so fast (about 100 fm/c, see Fig.S3 and Fig.S4 in the Supplement) that the non-localization of wave functions is kept during the separation. 
Compared to dynamical calculations, static constrained calculations~\cite{density-c,qy2} as described in {\textsf{Methods}},  have smaller entanglement after scission.
The static deformation-constrained calculations lead to almost zero entanglement after separation. 
The entanglement in static density-constrained calculations also approaches 
that of deformation-constrained calculations after separation.

To understand the connection between the sawtooth structures and quantum entanglement, the projection results of the single event around the peak are shown in Fig.\ref{FIG3}(c, d).
In Fig.\ref{FIG3}(c), it can be seen that in the final phase of scission, the light fragment acquires more excitation energy than the heavy fragment,
emphasizing the role of splitting dynamics in the energy sharing.

The positive slope in excitation energies is actually because of entanglement between two fragments. 
In Fig.\ref{FIG3}(e), the changes of projected binding energies due to changes of particle numbers 
in heavy fragments 
 are small as a function of mass number, either for adding or removing particles (similar situation in light fragments, see Fig. S5 in the Supplement).
Or say that the changes in particle numbers in one fragment become less costly by entangling with another fragment. 
Then the subtraction between projected binding energies in splitting and ground state energies results in a positive slope in excitation energies in Fig.\ref{FIG3}(c).

For comparison, disentangled two fragments are also studied with separated
density-constrained calculations of each fragment.
In Fig.\ref{FIG3}(c) distributions of  excitation energies of two disentangled fragments at $N_c$=1.0 after individual projections are parabolic curves. 
 The separated density-constrained calculations indicate that the positive sawtooth slope  of excitation energies is not due to deformation effects of fragments.
The fluctuated events without projections are also shown in Fig.\ref{FIG3}(c, d). It can be seen that their distributions are
rather flat associated with insufficient widths.
The slopes of excitation energies at  $N_c$=1.0 from different calculations are shown in Fig.\ref{FIG3}(f).
It can be seen that smaller entanglement in Fig.\ref{FIG3}(a) correspond to smaller slopes in Fig.\ref{FIG3}(f).
The deformation-constrained results with nearly zero entanglement also shows parabolic curves, which 
are similar to the disentangled results in Fig.\ref{FIG3}(c).
Therefore, based on  Fig.\ref{FIG3}(e) and  the above-mentioned comparison, we demonstrated that the dynamical entanglement is essential for the appearance of the positive sawtooth slopes in excitation energies and thus neutron multiplicities.

However, the slopes of $\langle N \rangle /Z$ ratios are not dependent on the neck size as shown in Fig.\ref{FIG3}(d).
The $\langle N \rangle /Z$ ratios are almost stable after $N_c$=2.0.
This can be understood that the partition of particles is earlier than the partition of energies between fragments~\cite{bulgac-energy}.
For disentangled fragments,
the $\langle N \rangle /Z$ ratios also don't change. 
Actually the projected 2D distributions of two disentangled fragments are distinctly asymmetric (see Fig.S6 in the Supplement).
Hence the quantum entanglement does play an important role in the sharing of particles but is not related to sawtooth structures in $\langle N \rangle /Z$ ratios.

\section*{Discussion}
Nuclear quantum features would fade away due to the thermalization in the dissipative fission process. 
This work finds that,  at low-energy fission, 
the dynamical quantum entanglement plays an essential role in the appearance of positive sawtooth slopes in excitation energies
and thus neutron multiplicities, but not in negative sawtooth slopes in neutron excess of fragments.
In both cases, the projection resulted sawtooth structures have steep slopes, which 
can be alleviated by random fluctuations. 
 When the statistical sharing mechanism is stressed,
the slope of the sawtooth structure is slightly underestimated~\cite{neutron3}.
 The sawtooth structures
would be washed out at high excitations when fluctuations are dominant, as demonstrated in experiments~\cite{neutron1,nzratio1,nzratio2}.
Recently the sawtooth structure in average angular momentum of fragments
has attracted great interests~\cite{bulgacspin,angular2,angular3}, which is not studied presently and is a later phase after neutron emissions.

Nuclear fission involves a few hundreds of strongly interacting  particles and
presents one of a kind quantum entanglement, in contrast to the entanglement of multiple photons
that are susceptible to external noises. 
The quantum entanglement in nuclear fission confronts with
 intrinsic flucutations due to higher-order dynamical correlations.
Our findings should be valuable for understanding the nature of quantum entanglement more broadly in mesoscopic systems.
Future experimental  event-by-event measurements of multiple fission observables are highly expected to be informative
to  extract the entanglement between two fragments.

\section*{Methods}\label{sec11}
\noindent{\bf Dynamical fission evolutions}\\
The theoretical framework used to describe the dynamical fission evolution beyond the saddle point in this work is
the time-dependent Hartree-Fock+BCS (TD-BCS) approach~\cite{TDBCS1,TDBCS2}.
Dynamical pairing correlations are very important for fission and are treated by the BCS method as opposed to the Bogoliubov method which induces further correlations but is computationally very expensive.
The TD-BCS equation is solved using the modified Sky3D code~\cite{sky3d}.
For the nuclear force, the SkM$^{*}$ parameterization is adopted~\cite{skm}, which is widely used in nuclear fission studies. 
 Besides, UNEDF1~\cite{UNEDF} force is also suitable for describing fission barriers, 
but is not expected to change dynamical evolutions significantly according to previous comparisons~\cite{marko}. 
The form of the pairing interaction
adopts the mixed density dependent pairing~\cite{mix-pair}. Details of TD-BCS calculations are given in our previous work~\cite{qy1,qy2}.

\noindent{\bf Double projection on particle numbers}\\
Based on TD-BCS solutions, the particle numbers of the fragments and the fissioning nucleus are
not well defined,
and they could be non-integer numbers. In this work, the double PNP
on the total space and a partial space is applied to determine the particle numbers of two complementary fragments.
 The projected states with particle numbers deviating from the
average number up to 10 particles are calculated.
The double projection operator is written as:
\begin{equation}
\begin{array}{ll}
\hat{P}^{q}(N^{q}_{T},N^{q}_{P})=& \frac{1}{4\pi^2}\iint d\theta_{T}d\theta_{P} \vspace{6pt}\\
&e^{i\theta_{T}(\hat{N}^{q}_{T}-N^{q}_{T})}e^{i\theta_{P}(\hat{N}^{q}_{P}-N^{q}_{P})} \\
\end{array}
\end{equation}
where $q$ denotes neutron or proton, $T/P$ denotes the projection on the total space or a partial space.
The particle number operator is: $\hat{N}^{q}_{P}=\int d\pmb{r} \hat{C}^{\dagger}(\pmb{r})\hat{C}(\pmb{r})\Theta(\pmb{r})$,
and $\Theta(\pmb{r})$ is a mask function to obtain an exclusive partial space.
The Gauss-Legendre integration method is used for the double integral and singularities in the projection kernel can be avoided.

The projection on each fragment configuration leads to a distribution of fragments, in which
the formation probability of each fragment is the expectation value of $\langle \Psi |\hat{P}^{q}(N^{q}_{T},N^{q}_{P})|\Psi\rangle$,
and is calculated by the integral over angles $\theta$ for $\langle\Psi|\hat{R}  (\theta_{T},\theta_{P})|\Psi\rangle=\langle\Psi|e^{i\theta_{T}\hat{N}_{T}}e^{i\theta_{P}\hat{N}_{P}}|\Psi\rangle$.
The projected  binding energy of each fragment is obtained by
\begin{equation}
E_{proj}=\frac{\langle\Psi|\hat{H}\hat{P}^{n}(N_{T},N_{P})\hat{P}^{p}(Z_{T},Z_{P})|\Psi\rangle}{\langle\Psi|\hat{P}^{n}(N_{T},N_{P})\hat{P}^{p}(Z_{T},Z_{P}|\Psi\rangle}
\label{proj}
\end{equation}
which is actually calculated as~\cite{egido},
\begin{equation}
\begin{aligned}
E_{proj}&=\!\!\int\!\!d\theta_{n}\!\!\int\!\!d\theta_{p} Y_{\theta_{n}}Y_{\theta_{p}}{\rm Tr}\{t(\rho^{n}_{\theta_{n}}+\rho^{p}_{\theta_{p}}) \\
&+\frac{1}{2}(\Gamma^{nn}_{\theta_{n}}\rho^{n}_{\theta_{n}} +\Gamma^{pp}_{\theta_{p}}\rho^{p}_{\theta_{p}}+\Gamma^{np}_{\theta_{p}}\rho^{n}_{\theta_{n}}+\Gamma^{pn}_{\theta_{n}}\rho^{p}_{\theta_{p}} \\
&-\Delta^{n}_{\theta_{n}}\overline{\kappa}^{n*}_{\theta_{n}}-\Delta^{p}_{\theta_{p}}\overline{\kappa}^{p*}_{\theta_{p}})\} \\
Y_{\theta}&=\langle\Psi|\hat{R}(\theta)|\Psi\rangle/\langle\Psi|\hat{P}_{N}|\Psi\rangle \\
\end{aligned}
\end{equation}
where $\rho_{\theta}$, $\kappa_{\theta}$ are transition densities. The excitation energy of each fragment is
obtained by the subtraction between the projected binding energy  during the splitting process and the ground state energy.
 Presently  excitation energies are calculated only when the denominator in Eq.(\ref{proj}) is larger than $5\times 10^{-4}$. 
This method has been applied to calculate excitation energies of products in multi-nucleon transfer reactions~\cite{sekizawa}.
For consistency, the ground state energies of fragments are obtained by Hartree-Fock+BCS plus PNP calculations. 
The one-body densities can also be obtained by ${\rho}(\vec{r})=\langle\Psi|\hat{\rho}\hat{P}_{N}|\Psi\rangle/\langle\Psi|\hat{P}_{N}|\Psi\rangle$~\cite{sheikh}.
In practical calculations, a series of transition densities like the current density, spin-orbit density have to be calculated at each $\theta$.
The calculations are very costly because the proton-neutron mixing terms such as $\Gamma^{np}$ and $\Gamma^{pn}$
involve fourfold integrations. In calculations of projected energies, to avoid heavy computational costs,
we applied two times of double PNP for each partial space.

\noindent{\bf Random fluctuations}\\
Random fluctuations are also included in TD-BCS as described in our previous work~\cite{qy1}.
The random transitions between single-particle levels are invoked at each step to mimic dynamical
fluctuations.
The occupation probabilities $\rho_i$ of randomly chosen levels around the Fermi surface
are modified by  $\delta\rho_i$. The maximum $\delta\rho_i$ is constrained by
the Pauli principle.
The transition probability between two levels is proportional to the Boltzmann factor $\rm exp(-|\Delta_E|/T_f)$ by considering their energy difference $\Delta_E$.
The paired levels should have exactly opposite changes in occupation probabilities.
 The effective temperature $T_{f}$ is a parameter of 1.5 MeV, which corresponds to a considerable spreading width of fission events in Ref.~\cite{qy1}.
 Note the choice of $T_{f}$ is not necessarily related to nuclear excitation energy as the fission yields from spontaneous fission
 have similar spreading widths.

\noindent{\bf Number of particles in neck}\\
\label{nc-m}
The neck size in the fission process is estimated by the number of particles
in the neck $N_c$.
\begin{equation}
N_c=\int dx \int dy  \int_{z_{neck}-\delta}^{z_{neck}+\delta} dz \rho(x,y,z)
\end{equation}
where $z_{neck}$ is the location of coordinate $z$ corresponding to the minimum of the nuclear density $\rho$($x,y,z$) in the neck,
$\delta$ is adopted as 0.8 fm. In this work $N_c$ is an indicator of the separation process.

\noindent{\bf Static constrained calculations}\\
The static density-constrained calculations are performed by iterative Hartree-Fock-BCS calculations,
to  reproduce the density distributions of
the fissioning nucleus in the evolution within TD-BCS calculations.
This approach has been applied to study nuclear fusion and fission before~\cite{density-c,qy2}.
Note that the pairing density is not constrained, which is dynamically changing fast in TD-DFT~\cite{bulgacprl,qy2}.

In separated density-constrained calculations, the density distribution of each
fragment is reproduced independently. In this case, fragments have same density distributions as in dynamical calculations, 
but
wave functions of two fragments
are disentangled. 

In  deformation-constrained calculations, the Hartree-Fock-BCS calculations
are performed to reproduce the quadrupole-octupole deformations ($\beta_2$, $\beta_3$) of the
fissioning nucleus within TD-BCS calculations. This is different from density-constrained calculations, which constrain
all degrees of deformations.

\noindent{\bf Entanglement magnitude}\\
In this work, we developed a method to estimate the entanglement magnitude between two fission fragments.
The entanglement is mainly caused by the non-localization of wave functions in two parts. 
The reduced entropy in the bipartite system in the Hartree-Fock approach can be strictly calculated as an estimation of entanglement magnitude~\cite{entropy}.
The single-particle wave functions can be projected into the subspace A and B respectively, using the projection operators $P_A$ and $P_B$. 
$P_A$ works as a projection to the subspace A by 
\begin{equation}
  P_A \phi_i (x,y,z)=\Theta(z-z_{neck}) \phi_i (x,y,z).
\end{equation}
where $\Theta$ is the step function. 
The overlap matrix in the subspaces can be diagonalized with the same unitary transformation~\cite{entropy}
\begin{equation}
\begin{aligned}
M(A)_{ij}&=\langle P_{A}\phi_{j},P_{A}\phi_{i}\rangle=S^{\dagger}diag(d_{i})S \\
M(B)_{ij}&=\langle P_{B}\phi_{j},P_{B}\phi_{i}\rangle=S^{\dagger}diag(1-d_{i})S      
\end{aligned}
\end{equation}
The single-particle operator can be rewritten as
\begin{equation}
\begin{aligned}
    a_{i}^{\dagger} & =\sum_{k}S_{ik}(\sqrt{d_{k}}A_{k}^{\dagger}+\sqrt{1-d_{k}}B_{k}^{\dagger}) \\
    a_{i} & =\sum_{k}S_{ik}^{*}(\sqrt{d_{k}}A_{k}+\sqrt{1-d_{k}}B_{k})
\end{aligned}
\end{equation}
where the operators  $A_{k}^{\dagger}$ and $B_{k}^{\dagger}$ only apply in the corresponding subspace.

In the TD-BCS approach,  wave functions after the Bogoliubov transformation can be written as a Slater determinant in the quasiparticle basis $\beta_{i}$.
\begin{equation}
\begin{aligned}
    \left(
\begin{array}{cc}
     \beta_{i} \\
     \beta_{i}^{\dagger} 
\end{array} 
\right) &= 
\left(
\begin{array}{cc}
     U_{c}S^{*} & V^{*}_{c}S  \\
     V_{c}S^{*}  &   U^{*}_{c}S  
\end{array}
\right)
\left(
\begin{array}{cc}
     \sqrt{d_{i}}A_{i}+\sqrt{1-d_{i}}B_{i}  \\
     \sqrt{d_{i}}A_{i}^{\dagger}+\sqrt{1-d_{i}}B_{i}^{\dagger} 
\end{array}
\right) 
\end{aligned}
\end{equation}

Next new quasiparticle basis $\tau_{i},\eta_{i}$ in the subspace $A$ and $B$ can be defined as:
\begin{equation}
\begin{aligned}
\left(
\begin{array}{cc}
     \tau_{i}  \\
     \tau_{i}^{\dagger} 
\end{array}
\right)
&=\left(
\begin{array}{cc}
     U' & V'^{*}  \\
     V'  & U'^{*}  
\end{array}
\right)
\left(
\begin{array}{cc}
    diag(\sqrt{d_{i}} ) A_{i}  \\
    diag(\sqrt{d_{i}}) A_{i}^{\dagger} 
\end{array}
\right) \\
\left(
\begin{array}{cc}
     \eta_{i}  \\
     \eta_{i}^{\dagger} 
\end{array}
\right)
&=\left(
\begin{array}{cc}
     U' & V'^{*}  \\
     V'  & U'^{*}  
\end{array}
\right)
\left(
\begin{array}{cc}
     diag(\sqrt{1-d_{i}}) B_{i}  \\
   diag(\sqrt{1-d_{i}})  B_{i}^{\dagger} 
\end{array}
\right) 
\end{aligned}
\end{equation}
where $ U' =  U_{c}S^{*} $ and $V' =V_{c}S^{*} $.
The anti-commutation relation of $\tau_{i},\eta_{i}$ can be derived as,
\begin{equation}
\begin{aligned}
\{\tau_{i},\tau_{j}^{\dagger} \} & = (U'{\rm diag}(d)U'^{\dagger}+V'^{*}{\rm diag}(d)V'^{T})_{ij}=O_{ij} \\
\{\eta_{i},\eta_{j}^{\dagger}\}&=(U'{\rm diag}(1-d)U'^{\dagger}+V'^{*}{\rm diag}(1-d)V'^{T})_{ij}\\
\end{aligned}
\end{equation}
It can be find that the overlap matrix in the subspace quasiparticle basis as:
\begin{equation}
\begin{aligned}
M'(A) & =O=W{\rm diag}(G)W^{\dagger} \\
M'(B) & =I-O=W{\rm diag}(1-G)W^{\dagger} \\
\end{aligned}
\end{equation}
The overlap matrix can be further diagonalized with the transformation $W$ to facilitate the calculations of entropy.
The density matrix can be expressed as a series of matrix direct product, which is similar to the Hartree-Fock state~\cite{entropy}.
The reduced  entanglement entropy $S_E$ is derived by partial tracing of the B subspace of the density matrix as
\begin{equation}
\begin{aligned}
\begin{array}{ll}
S_E  = & -{\rm{Tr}}[M'(A){\rm{ln}}(M'(A)) \\
 & + (I-M'(A)){\rm{ln}}(I-M'(A))] \\
 \end{array} \\
M'(A) = (UM(A)U^{\dagger}+V^{*}M(A)^{*}V^{T})
\end{aligned}
\end{equation}

In addition, the straightforward estimation of the number of entangled particles are also calculated for comparison.
The entanglement estimation is defined as 
\begin{equation}
\eta=\sum_i v_i^2 {\rm min}(P_i, 1-P_i)
\end{equation}
 where $v_i^2$ is the occupation probability of single particle levels. 
The non-localized wave function of each level has a distribution probability $P_i$ in the light fragment and $1-P_i$ in the heavy fragment. 
Then $\eta$ becomes zero if the wave function is localized in one fragment. Note that such an estimation is not strict and is not invariant under unitary
transformation. Nevertheless the trends of $\eta$ are very similar to the entanglement entropy.

\backmatter


\bmhead{Acknowledgments}
We are grateful to discussions with W. Nazarewicz, P. Stevenson, and F.R. Xu.
This work was supported by the National Key R$\&$D Program of China (Grant No. 2023YFA1606403, 2023YFE0101500)
and the National Natural Science Foundation of China under Grants
No.  11975032, 8200907308, 12335007,  and 11961141003.
This work has also been supported by the U.S. Department of Energy under award Nos. DE-SC0013365 (Office of Science) and DE-SC0023175 (NUCLEI SciDAC-5 collaboration).
We also acknowledge the funding support from the State Key Laboratory of Nuclear Physics and Technology, Peking University (No. NPT2023ZX01).
Computations in this work were performed in Tianhe-1A
located in Tianjin.

\section*{Declarations}
\begin{itemize}
\item Conflict of interest \\ The authors declare no competing interests.
\item Availability of data and materials \\
All relevant data from calculations are available online.
\item Code availability \\ 
The computer code used in this work is based on the Sky3D solver that is
available from ref.~\cite{sky3d}. The modifications of the basic code are made as described
in Methods. 
\item Authors' contributions \\
Y.Q. wrote the code and performed theoretical calculations of this study;  J.P. conceived
the project and wrote the manuscript; K.G. revised the manuscript and participated in discussions. 
\item Correspondence and requests for materials should be addressed to
Junchen Pei
\end{itemize}



\end{document}